\documentclass[onecolumn,preprint,showpacs,graphicx]{revtex4}
\usepackage{amssymb}
\usepackage{amsmath}
\usepackage{graphicx}

\begin{document}

\title{Molecular dynamics simulations of oxide memristors: crystal field
effects}
\author{S. E. Savel'ev$^{1}$, A. S. Alexandrov$^{1,2}$, A. M. Bratkovsky$%
^{2} $, and R. Stanley Williams$^{2}$}
\affiliation{$^1$Department of Physics, Loughborough University, Loughborough LE11 3TU,
United Kingdom\\
$^2$Hewlett-Packard Laboratories, 1501 Page Mill Road, Palo Alto, California
94304}

\begin{abstract}
We present molecular-dynamic simulations of memory resistors (memristors)
including the crystal field effects on mobile ionic species such as oxygen
vacancies appearing during operation of the device. Vacancy distributions
show different patterns depending on the ratio of a spatial period of the
crystal field to a characteristic radius of the vacancy-vacancy interaction.
There are signatures of the orientational order and of spatial voids in the
vacancy distributions for some crystal field potentials. The crystal field
stabilizes the patterns after they are formed, resulting in a non-volatile
switching of the simulated devices.
\end{abstract}

\maketitle

Electrical switching in thin-film oxide nanodevices attracts renewed
attention potentially enabling scaling of logic and memory circuits well
beyond the current state of the art techniques \cite{stan0}. The microscopic
nature of resistance switching and a charge transport in such devices is
still a subject of debate, but there is growing evidence that nonvolatile
switching requires some sort of atomic rearrangement of mobile ionic species
that drastically affects the current. The microscopic understanding of the
atomic-scale mechanism and identification of the material changes within the
device appears to be invaluable for controlling and improving the memristor
performance.

The number of oxygen vacancies within the volume 10$\times $10$\times $2 nm$%
^{3}$ of perspective nano-memristors, such as based on an amorphous layer of
titanium dioxide, TiO$_{2-x}$, could be as small as about a thousand, so
that the conventional statistical (diffusion) approach for dealing with
many-particle systems may fail. Recently, we have proposed a model for the
kinetic behavior of oxide memristors and studied it using the Molecular
Dynamics (MD) simulations of the Langevin equations describing the thermal
diffusion and drift of interacting oxygen vacancies \cite{saaw}. Our MD
simulations revealed a significant departure of the vacancy distributions
across the device from that expected within a standard drift-diffusion
approximation. The channel formation in systems like TiO$_{2}$, NiO, VO$_{2}$
is certainly accompanied by local heating that is witnessed by the local
emergence of high-temperature phases and observed by thermal microscopy and
other techniques \cite{strachan10,borg09,noh3w08}. Accounting for local
heating, we extended our MD modeling of oxygen vacancies driven by an
external bias voltage by taking into account temperature gradients in thin
films of oxide memristors \cite{saaw}. Temperature gradients, producing
strong variations of the rate of diffusion, affect the vacancy patterns in
memristors. Our simulations indicated that variations of temperature across
the memristor favor the formation of short-circuiting or shunting vacancy
channels. At the same time, considerable temperature gradients along the
sample can, by contrast, produce vacancy-poor regions where vacancy shunts
are not formed and electron percolation paths are blocked. Here we extend
our MD simulations of the memristor including the crystal field in the
Langevin equations.

\begin{figure}[tbp]
\begin{center}
\includegraphics[angle=-00,width=0.80\textwidth]{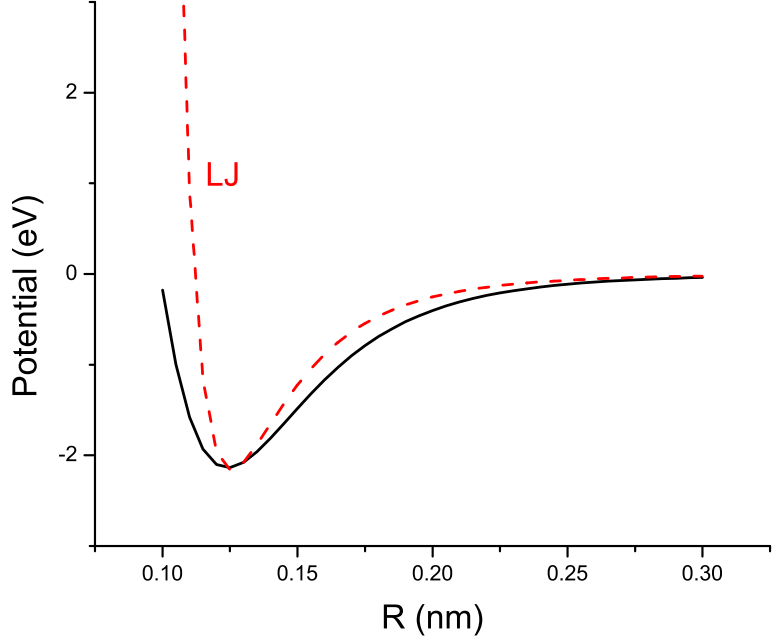}
\end{center}
\caption{(Color online) The Lennard-Jones potential (steeper curve) compared
with the O$^{2-}$ vacancy-vacancy interaction potential, Eq.(\protect\ref%
{pot}).}
\label{fig:pot}
\end{figure}

\begin{figure}[tbp]
\begin{center}
\includegraphics[angle=90,width=0.80\textwidth]{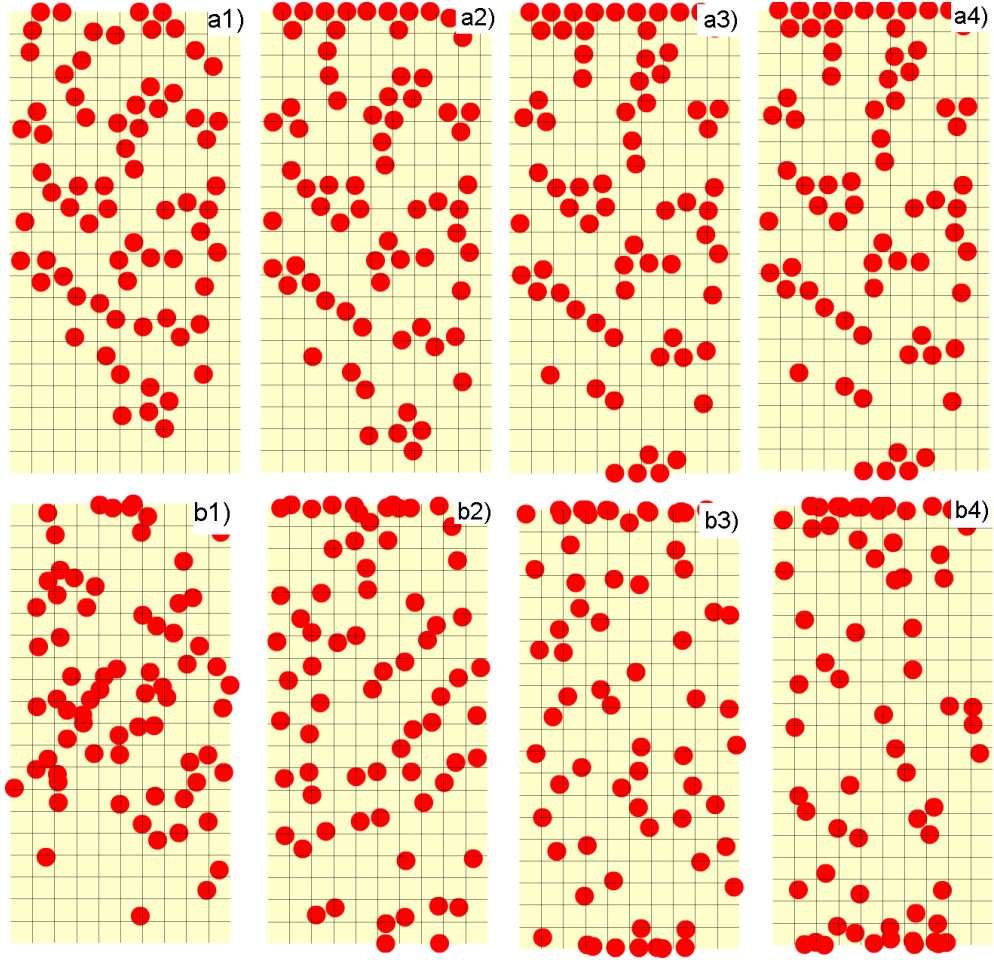}
\end{center}
\caption{(Color online) Vacancy dynamics in different crystal fields (upper
panels (a): $l=0.2$, lower panels (b): $l=2$) at different times after the
electric pulse has been applied: $\protect\tau=\protect\tau_0$ [a(1),b(1)], $%
\protect\tau=2\protect\tau_0$ [a(2),b(2)], $\protect\tau=3\protect\tau_0$
[a(3),b(3)], and $\protect\tau=4\protect\tau_0$ [a(4),b(4)]. }
\label{fig:1puls}
\end{figure}

Observations of oxygen vacancy migration and clustering in bulk \cite{miy}
and nanoscale \cite{kwon10,strachan10} samples of TiO$_{2}$, induced by an
electric-field, allows us to model a memristor \cite{saaw} with the vacancy
interaction potential, Fig.~\ref{fig:pot}, with the results shown in Fig.~%
\ref{fig:1puls}. In the model, there is a reduced rutile thin layer TiO$%
_{2-x}$ near one of the metallic electrodes stabilized by the Coulomb mirror
potential. Vacancies from this layer can drift toward the opposite electrode
pushed by a pulse of an electric field. Vacancies interact with each other
via the pairwise potential $W$, and with the electric field corresponding to
the time-dependent deterministic force $F$. The environment exerts an
independent Gaussian random force, $\vec{\xi}$ on each particle with zero
mean and intensity controlled by the temperature. Different from our
previous studies \cite{saaw}, we now include the periodic crystal field
potential $U(\boldsymbol{x}_{i})$ in the overdamped Langevin equations
describing the drift-diffusion of the i-th particle as
\begin{equation}
\eta {\frac{dx_{i}^{\alpha }}{{dt}}}=F_{i}^{\alpha }(\boldsymbol{x}%
_{i},t)-\sum_{i}{\frac{\partial U(\boldsymbol{x}_{i})}{{\partial }x{%
_{i}^{\alpha }}}}-\sum_{j\neq i}{\frac{\partial W(\boldsymbol{x}_{i}\mathbf{-%
}\boldsymbol{x}_{j})}{{\partial }x{_{i}^{\alpha }}}}+\sqrt{2k_{B}T\eta }\xi
_{i}^{\alpha }(t).  \label{lan}
\end{equation}%
where $x_{i}^{\alpha }$ is the $\alpha $-coordinate of the $i$th vacancy ($%
\alpha =x,y,z$), $\eta $ is the friction coefficient, $F_{i}^{\alpha }$ is
the $\alpha $-component of electric pulse force, and $\xi ^{i}$ is a random
force.

The O$^{2-}$ vacancy-vacancy interaction potential, which is responsible for
the clustering, is often modeled as \cite{meis}
\begin{equation}
W(R)=A\exp (-R/a_{0})-B/R^{6}+e^{2}/(\pi \epsilon _{0}\epsilon R),
\label{pot}
\end{equation}%
where the short-range repulsive and attractive parts are represented with
the parameters $A=22764.0$ eV, $a_{0}=0.01490$ nm, and $B=27.88\times 10^{-6}
$ eV nm$^{6}$ in TiO$_{2-x}$ \cite{rad}, and the long-range Coulomb
repulsion is given by the last term, see Fig.\ref{fig:pot}. Since we
simulate a limited number of vacancies, one can refer to each particle in
our simulations as a cluster of vacancies, where a cluster-cluster
interaction is more conveniently described by the combination of the
Lennard-Jones and Coulomb potentials acting with the force
\begin{equation}
F(R)=\frac{1}{R}\left\{
12E_{LJ}[(R_{min}/R)^{12}-(R_{min}/R)^{6}]+E_{c}R_{min}/R\right\} ,
\label{force}
\end{equation}%
where the relative strength of the Coulomb potential is given by $%
E_{c}/E_{LJ}=2$. This results in the position of the potential maxima $%
R_{max}\approx 2R_{min}$ and the height of the potential barrier on the
order of the depth of the potential well. The electric pulse strength is
taken such that the Coulomb force is about ten times stronger than the
maximum attracting force between vacancies, and the interactions are cut-off
at a distance of about $R_{min}/20$. The Lennard-Jones potential provides
somewhat \emph{stiffer} repulsion at small distances, but otherwise it
fairly fits Eq.(\ref{pot}) with the parameters $E_{LJ}\approx 2.16$ eV and $%
R_{min}\approx 0.125$ nm, Fig.\ref{fig:pot}.

The vacancy distribution depends on the boundary conditions, the sample
size, temperature gradients \cite{saaw}, and, in the present simulations, on
the ratio of the lattice constant, $a$ and the characteristic scale of the
interaction potential $R_{min}$. Hence, it is most convenient to measure
coordinates in units of $R_{min}$ by introducing $\mathbf{r}=\mathbf{x}%
/R_{min}$ while we measure time in units of $t_{0}=R_{min}^{2}/D$ by
introducing $\tau =t/t_{0}$, where $D=k_{B}T/\eta $ is the diffusion
coefficient. The Langevin equations in this dimensionless $(\mathbf{r},\tau )
$ space takes the following form:
\begin{equation}
{\frac{dr_{i}^{\alpha }}{{d\tau }}}=-{\frac{\partial \lbrack V(\boldsymbol{r}%
_{i})+U(\boldsymbol{r}_{i})]/(k_{B}T)}{{\partial }r{_{i}^{\alpha }}}}%
-\sum_{j\neq i}{\frac{\partial W(\boldsymbol{r}_{i}\mathbf{-}\boldsymbol{r}%
_{j})/(k_{B}T)}{{\partial }r{_{i}^{\alpha }}}}+\sqrt{2}\xi _{i}^{\alpha
}(\tau ),  \label{dimlan}
\end{equation}%
where $V(\boldsymbol{r}_{i})$ is the electric-field potential, and the
components of the (dimensionless) random force, $\xi _{i}^{\alpha }$,
satisfies the fluctuation - dissipation relation $\langle \xi _{i}^{\alpha
}(0)\xi _{j}^{\beta }(\tau )\rangle =\delta (\tau )\delta _{\alpha ,\beta
}\delta _{i,j}$. While the diffusion in TiO$_{2}$ is anisotropic, for
simplicity sake here and below we consider the uniform temperature across
the device and the isotropic diffusion of vacancies in a two dimensional
(2D) square lattice with the crystal field,
\begin{equation}
U(\mathbf{r})=U_{0}\sin (2\pi x/l)\sin (2\pi y/l),  \label{field}
\end{equation}%
where ${x,y}$ are two components of $\mathbf{r}$, and $l=a/R_{min}$.

We simulate Eqs.(\ref{dimlan}) for $N=70$ vacancies with $l=2$, $%
U_{0}=4E_{LJ}/3$ and with $l=0.2$, $U_{0}=E_{LJ}/6$, respectively, so that
the forces ($\sim U_{0}/a$) are of the same order of magnitude in both
cases. We place all the vacancies randomly near the bottom of a toy sample
and then let them evolve according to Eq.~(\ref{dimlan}) inside a
rectangular box, which mimics the actual sample. We use the 2D simulation
area with the aspect ratio $L_{y}/L_{x}=2$ and periodic boundary conditions
(BC) along the $x-$direction. Note that the periodic BC allow us to simulate
an infinite area sample using a rather small number of particles. To use
periodic BC, we include periodic images of the vacancies with respect to
vertical boundaries of the simulation box. We also incorporate opposite
polarity charges by adding mirror images of vacancies with respect to the
top and the bottom of the sample.

The results of simulations with the \emph{homogeneous} temperature [$%
k_{B}T\approx 1/30$ of the interaction potential well, Fig.\ref{fig:pot}],
are presented in Fig.~\ref{fig:1puls}. Different columns refer to different
moments in time [time intervals are shown relative to the rectangular
electric pulse duration, $\tau_{0}$]. The magnitude of the dimensionless
electric pulse force is taken as $L_y/2\tau_0$, so that vacancies are pushed
by the pulse to the center of the sample. The distribution is spread out and
its maximum moves towards the top of the sample with a constant velocity.

The important effect of the crystal field is that clusters appear more
ordered than those without the crystal field \cite{saaw}. Namely, there are
signatures of an orientational order at the angle of $45^{%
{{}^\circ}%
}$ for $l=0.2$, and of spatial voids in the vacancy distribution on the
order of a few lattice constants for $l=2$. This indicates that a large
variety of different vacancy phases (from liquid-like via smectic or nematic
liquid crystal-like to crystal-like phases) can be achieved in proper
domains of parameters (such as temperature, vacancy concentration, vacancy
pinning, electric pulse intensity, sample shape etc.) in this strongly
interacting system with competing thermal and quenched disorders. Actually,
our MD simulations may shed a light on what kind of microscopic
distributions could be behind a so-called "random circuit breaker network"
behavior phenomenologically introduced in Ref. \cite{model} used to explain
both bipolar and unipolar resistance switching in the Pt/SrTiO$_{x}$/Pt
capacitor and changeover between the two regimes.

Here, it is worth mentioning an analogy between vacancies in memristors and
superconducting vortices,  since in both cases long-range interactions,
which can be both repulsive and attractive, and thermal/quenched disorder
play a crucial role. It is well known \cite{blatter} that a huge variety of
vortex phases exists in superconductors. In addition to the standard vortex
crystal and vortex liquid phases, different types of smectic phases were
predicted and observed in superconductors. For anisotropic superconductors,
where the interaction between vortices can change from repulsive to
attractive at short distances, vortex chains were predicted and observed
that are reminiscent of the simulated filament-like clusters of vacancies in
memristors. Moreover, vortex avalanches can form due to a competition
between thermal/quenched disorder and vortex-vortex interactions that again
has a deep analogy with the observed vacancy clusters in memristors.
\begin{figure}[tbp]
\begin{center}
\includegraphics[angle=00,width=0.90\textwidth]{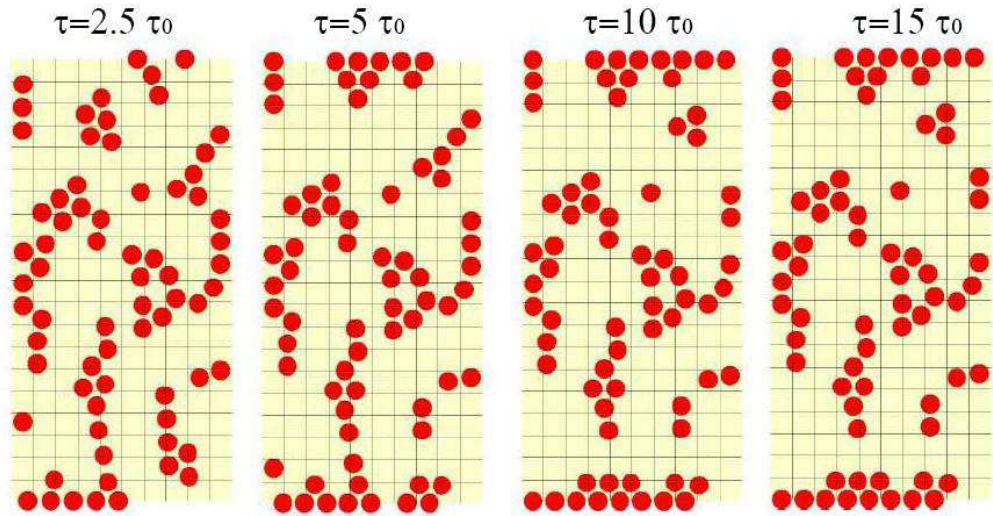}
\end{center}
\caption{(Color online) Vacancy dynamics in the crystal field with $l=0.2$
and with sufficiently low temperature after pulse is over. The evolution of
vacancies show a very slow (virtually none) dynamics well after the electric
pulse has been applied since the crystal field barriers are high with
respect to the thermal noise.}
\label{fig:3puls}
\end{figure}

Finally, it is well known that in addition to the equilibrium vortex phases,
many different metastable phases can occur when vortices are driven by a
magnetic field and relaxation between different phases is very slow
(logarithmic-time dynamics). We observe similar behavior in our simulations.
Indeed, as observed in our earlier MDs \emph{without} the crystal field \cite%
{saaw}, the vacancy distribution front keeps moving towards the top of the
sample gradually forming filamentary clusters even after switching the
voltage pulse off. If the vacancies continue to move for long times after
the end of the pulse, it would be difficult to build a non-volatile device.
These earlier simulations might be relevant for other systems, like Li$^{+}$%
, Na$^{+}$ or even Ag$^{+}$ ions in Silicon or Germanium. In these systems,
it is known that the ions continue to move after the pulse is over.
Remarkably, we observe in the present MDs that the crystal field stabilizes
the clusters after they are formed by lowering the mobility of vacancies.
These non-equilibrium metastable vacancy states relax extremely slowly,
since they have to overcome crystal field barriers to relax to states having
lower energy. Thus, vacancies in memristors can be frozen into a metastable
phase for an extremely long time, in analogy to vortices staying in
non-equilibrium critical states for many years. As a result, we see a
non-volatile behavior where the vacancies freeze right after the pulse if
the crystal field is large enough, as illustrated in Fig. (\ref{fig:3puls})
representing the vacancy distributions on a long time scale.

We conclude that manipulating the vacancy dynamics with the crystal field
and the vacancy-vacancy interactions could provide a rather rich
distribution of phases. We observe a tendency towards crystallization of the
vacancies, in agreement with the experimental devices, where the TiO$_2$
material in the switching channels is crystalline \cite{strachan10}. Our MD
simulations indicate that competing with the randomizing stochastic forces,
some crystal field potentials provide the orientational ordering and
stabilization of electrically conducting channels in memristors that are
vital for their nonvolatile memory property. Further 3D MDs of real
memristors with the number of oxygen vacancies as large as 1000 and above
should provide a powerful approach for understanding and improving these
devices.


\end{document}